\documentclass[11pt,a4paper,oneside]{article}

\usepackage{float}
\usepackage{amsmath,amssymb}
\usepackage[english]{babel}
\usepackage[dvips]{graphicx}

\newcommand{\LT}   {\left}
\newcommand{\RT}   {\right}

\newcommand{\MeV}  {{\text{~MeV}}}
\newcommand{\GeV}  {{\text{~GeV}}}
\newcommand{\EtAl} {\emph{et~al.}}
\newcommand{\CN}   {{\tilde\chi^{\pm,0}}}

\begin{document}

\begin{figure}[!t]
    \begin{flushright}
	INFN-FE-11-99 \\
	hep-ph/9907219
    \end{flushright}
\end{figure}

\setlength{\normalbaselineskip} {1.2\baselineskip}
\normalbaselines

\title{Diminishing ``charginos nearly degenerate with the lightest
  neutralino'' slit using precision data}
\author{M.~Maltoni$^{a,b,c}$ and M.~I.~Vysotsky$^{c,d}$}
\date{}

\maketitle

\begin{center} \small
    ${}^a$ Dipartimento di Fisica, Universit\`a di Ferrara, I-44100,
    Ferrara, Italy \\ 
    ${}^b$ Istituto TeSRE/CNR, Via Gobetti 101, I-40129 Bologna, Italy \\
    ${}^c$ INFN, Sezione di Ferrara, I-44100, Ferrara, Italy \\
    ${}^d$ ITEP, Moscow, Russia
\end{center}

\bigskip

\begin{abstract}
    Though LEP~II direct searches still cannot exclude a chargino nearly
    degenerate with the lightest neutralino if their mass is only slightly
    above half of the $Z$ boson mass, it can be excluded indirectly
    analyzing precision data. In this particular limit simple analytical
    formulas for oblique electroweak radiative corrections are presented.
\end{abstract}

\section{Introduction}

LEP~II is very effective in bounding from below masses of charginos which,
if this is kinematically allowed, should be produced in a pair in $e^+ e^-
\to \chi^+ \chi^-$ annihilation. The present bounds are $m_{\tilde\chi^\pm}
\gtrsim 90\GeV$ for the higgsino-dominated case and $m_{\tilde\chi^\pm}
\gtrsim 70\GeV$ for the gaugino-dominated case if sneutrino is not too
heavy~\cite{DELPHI98,DELPHI99}. However, when the lightest chargino and
neutralino (the latter being the LSP) are almost degenerate in mass,
the charged decay products of the light chargino are very soft, and the
above quoted bounds are no longer valid. Special search for such light
charginos has been performed recently by DELPHI collaboration, and the case
of $\Delta M^\pm \equiv m_{\tilde\chi_1^\pm} - m_{\tilde\chi_1^0} \lesssim
100\MeV$ is now excluded~\cite{DELPHI99}. However, still in the case of
$\Delta M^\pm \sim 1\GeV$ LEP~II does not provide a lower bound and
charginos as light as $45\GeV$ are allowed (this bound comes from the
measurements of $Z$ decays at LEP~I and SLC). The case of almost degenerate
chargino and neutralino can be naturally realized in SUSY and the
possibilities to find such particles are discussed in
literature~\cite{Gunion99}.

In this letter we investigate the radiative corrections to $m_W$ and to $Z$
boson decay parameters generated by such almost degenerate particles.
When their masses are close to $m_Z/2$ radiative corrections are large and
they spoil the perfect description of experimental data by the Standard Model.
Due to the decoupling property of SUSY models, when $m_\CN \gg m_Z$ the
radiative corrections are power suppressed.

\section{Discussion}

In the simplest supersymmetric extensions of the Standard Model the
chargino-neutralino sector is defined by the numerical values of four
parameters: $M_1$, $M_2$, $\mu$ and $\tan\beta$. The case of nearly
degenerate lightest chargino and neutralino naturally arise when:
\begin{enumerate}
  \item\label{case_a} $M_2 \gg \mu$: in this case the particles of interest
    form an $SU(2)$ doublet of Dirac fermions, whose wave functions are
    dominated by higgsinos;
  \item\label{case_b} $\mu \gg M_2$: in this way we get an $SU(2)$ triplet
    of Majorana fermions, with the wave functions dominated by winos.
\end{enumerate}
In this way we get higgsino- and gaugino-dominated scenarios,
correspondingly. Let us start from case~(\ref{case_a}).

\subsection{Higgsino-dominated case}

The degenerate Dirac doublet produces the following corrections to the three
functions $V_i$ which determine the values of radiative corrections (for
definitions of functions $V_i$ and further details about electroweak
radiative corrections see~\cite{NORV99}):
\begin{align}
    \delta^{\tilde h} V_m &= \frac{16}{9} \LT[
       \LT( \frac{1}{2} - s^2 + s^4 \RT) \LT( 1 + 2\chi \RT) F(\chi) 
       - \LT( \frac{1}{2} - s^2 \RT) \LT( 1 + 2\frac{\chi}{c^2} \RT)
        F\LT( \frac{\chi}{c^2} \RT) - \frac{s^4}{3} \RT], \\
    \delta^{\tilde h} V_A &= \frac{16}{9} \LT( \frac{1}{2} - s^2 + s^4 \RT)
      \LT[ \frac{12 \chi^2 F(\chi) - 2\chi - 1}{4\chi - 1} \RT], \\
    \delta^{\tilde h} V_R &= \frac{16}{9} c^2 s^2 
      \LT[ \LT( 1 + 2\chi \RT) F(\chi) - \frac{1}{3} \RT],
\end{align}
where $\chi \equiv (m_\CN / m_Z)^2$, the function $F$ is defined in
Appendix~B of Ref.~\cite{NORV99}, and $s^2$ ($c^2$) is the sine (cosine)
squared of the electroweak mixing angle $\theta$.

\begin{figure}[H] \centering
    \includegraphics[width=\textwidth]{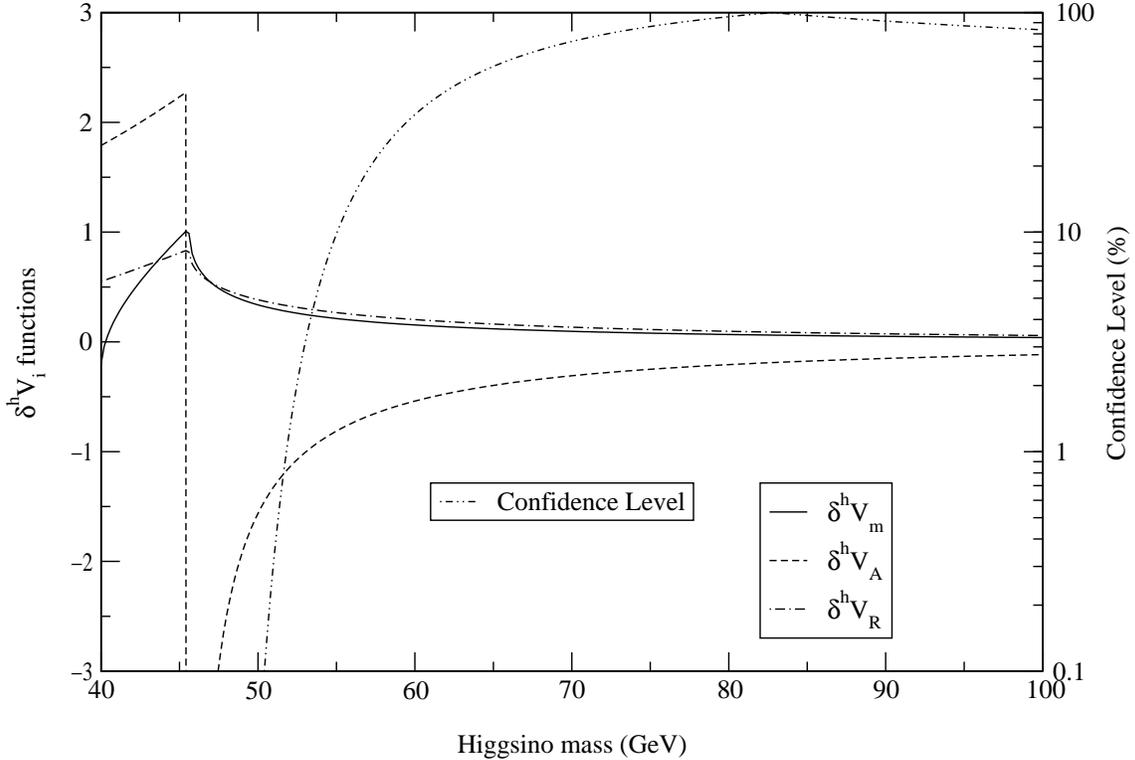}
    \caption{Dependence of the $\delta^{\tilde h} V_i$ functions (left
      Y-axis) and of the confidence level (right Y-axis) on the light gaugino
      mass $m_\CN$, in the limit $M_2 \gg \mu$ (higgsino-dominated case).}
    \label{fig10}
\end{figure}

Comparing the experimental data with Standard Model formulas~\cite{NORV99},
we obtain that the $\chi^2$ for the new physics contributions to $V_i$ can be
computed in the following way:
\begin{gather}
    \label{eq40a} \chi^2 = C_{ij} 
      \LT( \delta_\mathrm{NP} V_i - \overline{\delta V_i} \RT)
      \LT( \delta_\mathrm{NP} V_j - \overline{\delta V_j} \RT) \\[2mm]
    \label{eq40b} \begin{pmatrix}
	C_{mm} & C_{mA} & C_{mR} \\
	C_{mA} & C_{AA} & C_{AR} \\
	C_{mR} & C_{AR} & C_{RR}
    \end{pmatrix} = \begin{pmatrix}
	7.28   & 0      &  0     \\ 
	0      & 7.24   &  2.54  \\
	0      & 2.54   & 23.03
    \end{pmatrix}; \qquad \begin{pmatrix}
	\overline{\delta V_m} \\
	\overline{\delta V_A} \\
	\overline{\delta V_R}
    \end{pmatrix} = \begin{pmatrix}
	-0.07  \\ 
	-0.33  \\
	+0.01
    \end{pmatrix}.
\end{gather}
In Fig.~\ref{fig10} the functions $\delta^{\tilde h} V_i$ are plotted
against the chargino-neutralino mass $m_\CN$. Comparing this graph with
formulas~\eqref{eq40a} and~\eqref{eq40b}, we see that at $95\%$ C.L. the
bound $m_\CN \gtrsim 54\GeV$ should be satisfied. Note that the main
contribution to $\chi^2$ comes from $\delta^{\tilde h} V_A$, which is
singular at $m_\CN = m_Z/2$. 
This singularity is not physical and our formulas are valid only for $2
m_\CN \gtrsim m_Z + \Gamma_Z$; the existence of $\chi^\pm$ with a mass
closer to $m_Z/2$ will change Z-boson Breit-Wigner curve, therefore it is
also not allowed.
The importance of the $Z$ wave function renormalization for the case of
light charginos was emphasized in~\cite{Barbieri92}.

\vspace{\baselineskip}

\subsection{Wino-dominated case}

\begin{figure}[H] \centering
    \includegraphics[width=\textwidth]{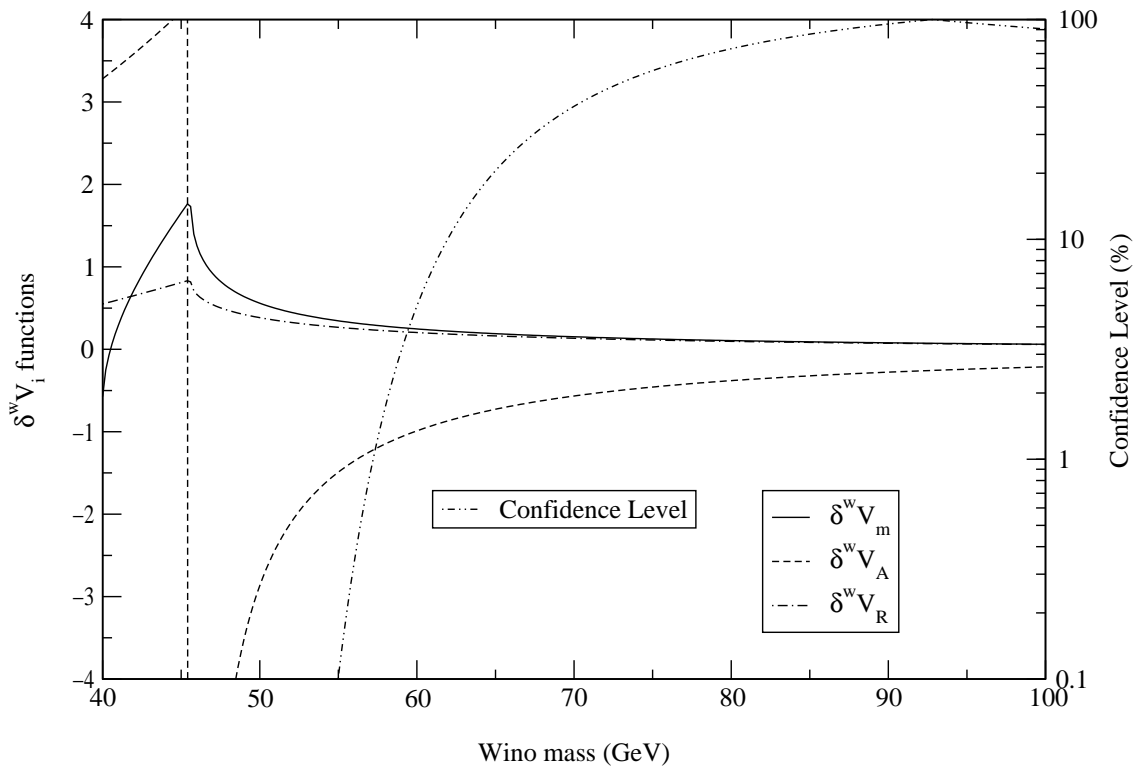}
    \caption{Dependence of the $\delta^{\tilde w} V_i$ functions (left
      Y-axis) and of the confidence level (right Y-axis) on the light gaugino
      mass $m_\CN$, in the limit $\mu \gg M_2$ (wino-dominated case).}
    \label{fig20}
\end{figure}

For case~(\ref{case_b}) (gaugino dominated states) the expressions for the
 $\delta^{\tilde w} V_i$ functions are:
\begin{align}
    \delta^{\tilde w} V_m &= \frac{16}{9} \LT[ c^4 \LT( 1 + 2\chi \RT) F(\chi)
      - \LT( 1 - 2s^2 \RT) \LT( 1 + 2\frac{\chi}{c^2} \RT)
        F\LT( \frac{\chi}{c^2} \RT) - \frac{s^4}{3} \RT], \\
    \delta^{\tilde w} V_A &= \frac{16}{9} c^4
      \LT[ \frac{12 \chi^2 F(\chi) - 2\chi - 1}{4\chi - 1} \RT], \\
    \delta^{\tilde w} V_R &= \frac{16}{9} c^2 s^2 
      \LT[ \LT( 1+2\chi \RT) F(\chi) - \frac{1}{3} \RT].
\end{align}
The values of these functions are shown in Fig.~\ref{fig20} and at 95\%
C.L. we get $m_\CN \gtrsim 60\GeV$.

Let us remark that, although this and the previous bounds have been obtained
respectively in the limits $|\mu| \to \infty$ and $|M_2| \to \infty$ (in
which case the mass splitting $\Delta M^\pm$ is exactly zero), we have
verified numerically using equations from Ref.~\cite{Hollik99} that they are
still valid for values of $|M_2|$ and $|\mu|$ small enough to allow for
$\Delta M^\pm \sim 1\GeV$.

\vspace{\baselineskip}

\section{Conclusions}

Since there are a number of new additional particles in SUSY extensions, we
will briefly discuss their contributions to the functions $V_i$. In the
considered limits the remaining charginos and neutralinos are very heavy, so
they simply decouple and produce negligible contributions.
The contributions of the three generations of sleptons (with masses larger
than $90\GeV$) into $V_A$ are smaller than $0.1$, so they can be safely
neglected.
The contributions of squarks of the first two generations are also negligible
since they should be heavier than Tevatron direct search bounds;
taking $m_{\tilde q} \gtrsim 200\GeV$, we have $| \delta^{\tilde q} V_i |
\lesssim 0.1$.
Concerning the contributions of the third generation squarks, they are
enhanced by the large top-bottom mass difference and are not negligible.
However, being positive and almost universal~\cite{Gaidaenko98}, they do not
affect our analysis: compensating negative contributions of
chargino-neutralino into $V_A$ they will generate positive contributions to
$V_R$ and $V_m$, and $\chi^2$ will not be better.
When squarks are heavy enough (for $m_{\tilde b} \gtrsim 300\GeV$), they
simply decouple and their contributions become negligible as well.

The last sector of the theory to be discussed is Higgs bosons. Unlike the
case of Standard Model now we have one extra charged higgs and two extra
neutral higgses. Their contributions to radiative corrections were
studied in detail in~\cite{Chankowski94}. According to Fig.~2 from that
paper it is clear that the contributions of MSSM higgses (and $SU(2) \times
U(1)$ gauge bosons) equal with very good accuracy those of the Standard
Model with the mass of SM higgs being equal to that of the lightest neutral
higgs in SUSY generalization. That is why the contributions from the
gauge-Higgs sector of the theory also cannot compensate those of the light
chargino-neutralino.

Apart from oblique corrections (those arising from vector bosons self
energies) which have been considered in this letter, there are process
dependent vertex and box corrections. However, due to LEP~II and Tevatron
low bounds on squarks and sleptons masses they are small.

\section{Acknowledgments}

We are grateful to F.~L.~Villante for useful discussion on $\chi^2$
analysis and to P.~H.~Chankowski for bringing Ref.~\cite{Barbieri92} to our
attention. Investigation of M.~V. are supported by the grants of RFBR
No.~98-07-90076 and 98-02-17372.


\begin{thebibliography}{99}

\bibitem{DELPHI98}
  {\rm DELPHI Coll., P.~Abreu \EtAl}:
  {\it CERN-EP/98-176}, {\rm (1998)};
 
  {\rm ALEPH Coll.}:
  {\it CERN-EP/99-014}, {\rm (1999)}.

\bibitem{DELPHI99}
  {\rm DELPHI Coll., P.~Abreu \EtAl}:
  {\it CERN-EP/99-037}, {\rm (1999)}.

\bibitem{Gunion99}
  {\rm J.~F.~Gunion, S.~Mrenna}:
  {\it UCD-99-01}, {\rm (1999)};

  {\rm C.~H.~Chen, M.~Drees, J.~F.~Gunion}:
  {\it Phys. Rev. Lett.} {\bf 76}, {\rm (1996) 2002};

  {\rm J.~L.~Feng \EtAl}:
  {\it IASSNS-HEP-99-19} {\rm (1999)}.

\bibitem{NORV99}
  {\rm V.~A.~Novikov \EtAl}:
  {\it hep-ph/9906465} {\rm (1999)}.

\bibitem{Barbieri92}
  {\rm R.~Barbieri \EtAl}:
  {\it Phys. Lett.} {\bf B279}, {\rm (1992) 169}.

\bibitem{Hollik99}
  {\rm W.~Hollik, C.~Schappacher}:
  {\it Nucl. Phys.} {\bf B545}, {\rm (1999) 98}.

\bibitem{Gaidaenko98}
  {\rm I.~V.~Gaidaenko \EtAl}:
  {\it JETP Lett.} {\bf 67}, {\rm (1998) 761}; {\it hep-ph/9812346}.

\bibitem{Chankowski94}
  {\rm P.~H.~Chankowski \EtAl}:
  {\it Nucl. Phys.} {\bf B417}, {\rm (1994) 101}.

\end{thebibliography}
\end{document}